\documentclass[11pt]{svjour2a}
\usepackage{graphics}
\newcommand{\D}{{\rm d}}

\begin{document}
\journalname{General relativity and gravitation}
\title{Bousso entropy bound in self-gravitating gas of massless particles}

\author{Jan Ger\v{s}l}
\institute{Institute of Physical Electronics, Faculty of Science, Masaryk University, Kotl\'{a}\v{r}sk\'{a} 2, 611 37 Brno, Czech Republic, 
\email{janger@physics.muni.cz}} 
\maketitle

\begin{abstract}
The Bousso entropy bound is investigated in a static spherically symmetric spacetime filled with an ideal gas of massless bosons 
or fermions. Especially lightsheets generated by spheres are considered. A statistical description of the gas is given. 
Conditions under which the Bousso bound can be violated are discussed and 
it is shown that a possible violating region cannot be arbitrarily large and that it is contained inside a sphere of unit Planck 
radius if the number of independent polarization states $g_s$ is small enough. It is also shown that the central temperature must 
exceed the Planck temperature in order to get a violation of the Bousso bound for $g_s$ not too large.
\end{abstract}
\section{Introduction}
The Bekenstein-Hawking formula for entropy of black holes combined with the generalised second law 
of thermodynamics \cite{Bekenstein} led to an idea that entropy of matter contained in a specified region of spacetime should be 
restricted by the area of a specified two dimensional spacelike surface \cite{Suss,W2,Bousso2,Mar} (for review on this topic see \cite{Bousso1}). 
R. Bousso formulated a hypothesis of this kind which should be valid in general spacetime in all situations where 
quantum effects are not of great importance \cite{Bousso2}. The Bousso entropy bound states the following.  
Let $(M,g)$ be a spacetime satisfying Einstein's equations and the dominant energy condition. Let $B$ be a two 
dimensional spacelike surface in $M$. Consider a null hypersurface $L$ (called lightsheet of $B$) formed by a congruence of null 
geodesics starting from $B$ orthogonally to it such that the expansion $\theta$ of the congruence is everywhere nonpositive and each geodesic 
is terminated if $\theta\rightarrow -\infty$, and otherwise is extended as far as possible. If we denote $S(L)$ the entropy of matter on the lightsheet and 
$A(B)$ the area of $B$, then the Bousso entropy bound states that $S(L)\leq A(B)/4$. (In all the text we use Planck units in which 
$c=1,G=1,h=2\pi ,k=1$, where $h$ is the nonreduced Planck constant and $k$ is the Boltzmann constant.) This bound was successfully 
tested in various situations \cite{Bousso1,Husain,Lemos}. Sufficient conditions for this bound were formulated in 
purely local terms by Flanagan, Marolf and Wald (FMW conditions) \cite{FMW}. 

The aim of this article is to investigate the Bousso bound in a special case of static spherically symmetric spacetime filled 
with an ideal gas of massless particles (radiation). For lightsheets generated by spheres the bound reduces to a statement that the entropy 
of the matter contained in a sphere does not exceed a quarter of the area of the sphere. 

In the first section a statistical description of bosonic or fermionic radiation in a stationary spacetime is given. In the second section 
some properties of solutions of the Einstein's equations for static spherically symmetric spacetime filled with radiation are discussed. 
In the third section the behaviour of the ratio of entropy contained in a sphere and quarter of area of the sphere across the spacetime is 
investigated. Finally the FMW conditions are discussed. 

\section{Statistical description of radiation}
In this section we apply some results of relativistic kinetic theory and statistical mechanics to obtain an energy-momentum tensor of radiation in a stationary spacetime and its entropy. The particles of the gas of radiation are treated as test particles moving in a mean gravitational field of the rest of 
the gas.  \\\\
{\it One particle phase space.} 
In a spacetime $(M,g)$ 
a classical state of a massless particle located at a spacetime point $x$ is described by a future directed null 
vector $p\in T_xM$. The set of all such vectors - a future lightcone in $T_xM$ - is denoted $L_x$. The one particle 
phase space at an instant of time represented by a spacelike hypersurface $\Sigma$ in spacetime 
is formed by the union of future lightcones at all the points of $\Sigma$. This phase space is denoted as $PS_\Sigma$.  
Given a coordinate system
\footnote{The Latin indices go from 1 to 3, the Greek ones go from 0 to 3.}
$x^i$ on $\Sigma$, we can construct coordinates $(x^i,p_i)$ on $PS_\Sigma$, where the momentum coordinates of a particle with fourvelocity $p$ are given by $p_i=g(\partial _i,p)$. Using these coordinates, the volume element on $PS_\Sigma$ can be expressed in a particularly simple manner (for details see e.g. \cite{Stew})
\begin{equation}{\rm vol}_{PS_\Sigma} =\D x^1\wedge\D x^2\wedge\D x^3\wedge\D p_1\wedge\D p_2\wedge\D p_3.
\label{volPS}
\end{equation} 
{\it Density of particles in phase space.} 
In semiclassical considerations a state of particle is given by a 6D-cube of volume $h^3$ in 
the phase space and by specification of quantities like polarization or spin. Considering a grandcanonical distribution 
of bosonic or fermionic ideal gas, 
the mean number of particles in certain state is 
given by Bose-Einstein or Fermi-Dirac statistics respectively. 
If we denote the number of independent polarizations $g_s$, the mean density of particles in phase space 
$\mathcal{N}$ (for details on the definition of $\mathcal{N}$ see e.g. \cite{MTW,Stew}) is then given by
\begin{equation} \mathcal{N}=\frac{g_s}{h^3}\left({\rm e}^{(H-\mu _0)/kT_0}+z\right)^{-1},\end{equation}
where $z=-1$ for bosons, $z=1$ for fermions, $T_0,\mu _0$ are constants related to the temperature and chemical potential, 
as we will see later. In a stationary spacetime with a timelike future oriented Killing field $\xi$, the Hamiltonian $H$ of 
a geodesically moving particle with fourmomentum $p$ is given by 
\begin{equation} H=-g(p,\xi). \label{Ham}\end{equation}
{\it Energy-momentum tensor.} Consider an orthonormal frame $e_\mu$ in $T_x M$ with $e_0$ 
future directed timelike. We can define 
coordinates $\bar{p}_i$ on $L_x$ by $\bar{p}_i =g(e_i ,p)$ and a function 
$\bar{p}_0=g(e_0,p)=-\sqrt{\delta ^{ij}\bar{p}_i\bar{p}_j}$ on $L_x$. 
The $3$-form 
\begin{equation}{\rm vol}_{L_x} =-\frac{1}{\bar{p}_0}\D \bar{p}_1\wedge\D \bar{p}_2\wedge\D \bar{p}_3\end{equation}  
represents a volume 
element on $L_x$ \cite{MTW,Stew}. The energy-momentum tensor of a system with density function $\mathcal{N}$ is 
given by \cite{MTW,Stew}
\begin{equation}T_{\mu\nu}(x) =\int _{L_x}{\cal N}p_\mu p_\nu{\rm vol}_{L_x}.\end{equation}
Denote $F=\sqrt{-\xi _\mu\xi ^\mu}$ and choose $e_0=\xi /F$. Then we have $H=-F\bar{p}_0=
F\sqrt{\delta ^{ij}\bar{p}_i\bar{p}_j}$. After some calculations we obtain the following 
nonzero components of $T_{\mu\nu}$ in the frame $e_\mu$  
\begin{eqnarray}
\rho &\equiv &T_{00}=4\pi\frac{g_s}{h^3}\left(\frac{kT_0}{F}\right)^4\int^\infty _0\D q\ q^3\left({\rm e}^{q-a}+z\right)^{-1} \label{rho}\\ 
P &\equiv &T_{ii}=\frac{1}{3}\rho ,\ \ \ \ \ i=1,2,3,\label{tlak}
\end{eqnarray}
where we have introduced a parameter $a=\mu _0 /kT_0$. For fermions $a\in(-\infty ,\infty)$, for bosons $a\in(-\infty ,0]$ in order to get nonnegative values of $\mathcal{N}$.
\footnote{In case of photonic black body radiation in equilibrium with walls of cavity $a$ is equal to zero.} 
Thus we have obtained the perfect fluid form of $T_{\mu\nu}$ with $P=\frac{1}{3}\rho$, as we expect for radiation. 
We see that $\rho$ and $P$ depend on 
the spacetime position only through $F$ and we can write $P=P_0F^{-4},\rho =3P_0F^{-4}$, 
where $P_0$ is a constant depending only on the constant parameters $P_0=P_0(T_0,a,g_s,z)$.\\\\
{\it Entropy.} The grandcanonical potential $\Omega _\Sigma$ of an ideal bosonic or fermionic gas 
on a spacelike hypersurface $\Sigma$ is given by 
\begin{equation}
\Omega _\Sigma =-zkT_0\frac{g_s}{h^3}\int_{PS_\Sigma}{\rm ln}\left(1+z\ {\rm e}^{-(H-\mu _0)/kT_0}\right){\rm vol}_{PS _\Sigma}.
\label{potential}
\end{equation} 
Entropy on $\Sigma$ then is $S_\Sigma =-\partial \Omega _\Sigma /\partial T_0$. To proceed with the integral 
(\ref{potential}), consider a field of orthonormal frames $e_\mu$ along $\Sigma$ with $e_0=\xi /F$ and 
a coordinate system $x^i$ on $\Sigma$. A frame $e_\mu(x)$ 
generates a coordinate system $\bar{p}_i$ on $L_x$ as described above. 
First we will express the volume form (\ref{volPS}) in terms of coordinates $(x^i,\bar{p}_i)$.
\begin{lemma} Suppose that $\partial /\partial x ^i=B_ie_0+A^j_ie_j$ with ${\rm det}A^j_i>0$ all over $\Sigma$. 
Let us denote $h_{\mu\nu}=g_{\mu\nu}+\xi _\mu\xi _\nu /F^2$ and $h_{ij}=h_{\mu\nu}(\partial /\partial x ^i)^\mu(\partial /\partial x ^j)^\nu$. 
Then the volume element (\ref{volPS}) on $PS_\Sigma$ has the form
\begin{equation}{\rm vol}_{PS _\Sigma} =\sqrt{{\rm det}h_{ij}}\ (1+\frac{C_i\bar{p}^i}{\bar{p}_0})\ 
\D x^1\wedge\D x^2\wedge\D x^3\wedge\D \bar{p}_1\wedge\D \bar{p}_2\wedge\D \bar{p}_3,
\label{volPS2}
\end{equation} 
where $C_i=B_j(A^{-1})^j_i$, $\bar{p}^i=\bar{p}_j\delta ^{ji}$ and $\bar{p}_0=-\sqrt{\delta ^{ij}\bar{p}_i\bar{p}_j}$. {\rm (The proof is given in appendix \ref{Z}.)}  
\end{lemma}
The term $(1+C_i\ \bar{p}^i/\bar{p}_0)$ in (\ref{volPS2}) splits the integral (\ref{potential}) into two terms. The term 
with $C_i\bar{p}^i$ will vanish, because the integrated function is odd in $\bar{p}_i$. 
Introducing spherical coordinates for the momentum part of the integral with $q=F\sqrt{\delta ^{ij}\bar{p}_i\bar{p}_j}/kT_0$, 
writing $4\pi q^2$ as $4\pi /3\  \D q^3/\D q$ and integrating by parts we get 
\begin{equation}
\Omega _\Sigma =-\int _\Sigma \D ^3x\sqrt{{\rm det}h_{ij}}\ FP,
\label{potential2}
\end{equation}
where $P$ is given by (\ref{tlak}). For the entropy of radiation 
$S_\Sigma =-\partial \Omega _\Sigma /\partial T_0$ we have 
\begin{equation}
S_\Sigma =\int _\Sigma \D ^3x\sqrt{{\rm det}h_{ij}}\ \sigma,
\label{entropy}
\end{equation}
where $\sigma =F\partial P/\partial T_0$ 
is a rest density of entropy. For $\sigma$ we get (remember that $a=\mu _0 /kT_0$)
\begin{equation}
\sigma =k\ 4\pi\frac{g_s}{h^3}\left(\frac{kT_0}{F}\right) ^3\int^\infty _0\D q\ q^2(4q/3-a)
\left({\rm e}^{q-a}+z\right)^{-1}.
\label{sigma}
\end{equation}
Again $\sigma$ depends on spacetime position only through $F$ and we can write $\sigma =\sigma _0F^{-3}$, where $\sigma _0$ is 
a function of constant parameters $\sigma _0=\sigma _0(T_0,a,g_s,z)$. 

Next we can define a notion of entropy on a lightlike hypersurface, which is needed in the Bousso entropy bound. For this we will use 
the following lemma.
\begin{lemma} Let $\Sigma _1,\Sigma _2$ be spacelike hypersurfaces. Suppose that a set of integral curves of $\xi$ 
which intersect $\Sigma _1$ is the same as an analogous set for $\Sigma _2$. Then $S_{\Sigma _1}=S_{\Sigma _2}$. {\rm (The proof is given in appendix \ref{A}.)}
\end{lemma}
\begin{definition} Let $L$ be a lightlike hypersurface and $\Sigma$ an arbitrary spacelike hypersurface satisfying the condition that the set of integral 
curves of $\xi$ intersecting $\Sigma$ is the same as the analogous set for $L$. It is natural to define $S_L=S_\Sigma$.
\end{definition}
{\it Mean number of particles.} The mean number of particles on $\Sigma$ is given by $N_\Sigma =-\partial\Omega _
\Sigma /\partial\mu _0 =\int_{PS_\Sigma}\mathcal{N}{\rm vol}_{PS_\Sigma}$. Following the same lines of argumentation as above 
we get $N_\Sigma =\int_\Sigma\D ^3x\sqrt{{\rm det}h_{ij}}\ n$, where the rest concentration of particles $n$ is given by 
\begin{equation}
n=4\pi\frac{g_s}{h^3}\left(\frac{kT_0}{F}\right)^3\int^\infty _0\D q\ q^2\left({\rm e}^{q-a}+z\right)^{-1}.
\label{n}
\end{equation}

Now we can interpret $T_0,\mu _0$ in terms of temperature and chemical potential. The first law of thermodynamics in terms of densities reads $\D \rho =T\D\sigma +\mu\D n$, where $T$ is the temperature and $\mu$ the chemical potential of the system. If $\rho ,\sigma$ and $n$ are treated as functions of $T_0/F$ and $a$, it can be shown by explicit calculation that $\D\rho =(T_0/F)\D\sigma +(\mu _0/F)\D n$. Thus we see that $T=T_0/F$ and $\mu =\mu _0/F$. This agrees with the result of Tolman \cite{Tolman}, which says that a product of $T$ and $F$ is constant across a static spacetime in thermodynamic equilibrium. 

\section{Einstein's equations}
In this paper we will focus on static spherically symmetric configurations of radiation with 
the energy-momentum tensor given in the previous section. Using the standard metric ansatz 
\begin{equation}
\D s^2=-F(r)^2\D t^2+\left(1-\frac{2m(r)}{r}\right)^{-1}\D r^2+r^2\D \Omega ^2,
\label{met1}
\end{equation} 
the Einstein's equations for functions $m(r),F(r)$ are \cite{Wald}
\begin{eqnarray}
m^\prime &=& 4\pi r^2\rho\label{ERm}\\
F^\prime &=&F\ \frac{m+4\pi r^3P}{r(r-2m)},
\label{ERF}
\end{eqnarray}
where $P=P_0F^{-4}$ and $\rho =3P$. 
Prime denotes derivative with respect to $r$. 
The third equation $P^\prime =-(P+\rho)F^\prime /F$, following from the equations of motion $\nabla _\mu T^{\mu\nu}=0$ 
is satisfied identically. 

A solution $F(r),m(r)$ of equations (\ref{ERm}), (\ref{ERF}) is specified by values of parameters 
$m(0),\ F(0) ,\ P_0$. We have to set $m(0)=0$ in order to get a solution regular in the centre. 
New values of parameters $F(0)_{new}=\alpha F(0)_{old},\ P_{0new}=\alpha ^4P_{0old}$ lead to a new solution 
$F_{new}(r)=\alpha F_{old}(r), m_{new}(r)= m_{old}(r)$. The new solution is isometric to the old one and physical quantities 
like $\rho(r), P(r)$ remain unchanged. We can use this gauge freedom to put $F(0)=1$ without loss of generality.

Let's choose a value $P_0^\ast$ of the parameter $P_0$, 
define $\alpha =P_0/P_0^\ast $ and denote 
$F^\ast(r),\ m^\ast(r)$ a solution for $\alpha =1$. 
A solution of (\ref{ERm}), (\ref{ERF}) is then specified by the value of $\alpha$ and has the form 
\begin{equation}
F(r;\alpha)=F^\ast(\sqrt{\alpha}r),\ \ \ \ m(r;\alpha)=m^\ast(\sqrt{\alpha}r)/\sqrt{\alpha}.
\label{param}
\end{equation}

The radiation in a static spherically symmetric spacetime has infinite extent. This can be seen in the following way. 
If the matter has finite extent, there is a radius 
$r_b$ where $P(r_b)=0$ and $P(r)>0$ for $r<r_b$. Since $2m(r)/r<1$ for $r\leq r_b$ (see theorem 2 in \cite{RenSch}), 
from $(\ref{ERF})$ we get that $({\rm ln}F(r))^\prime$ is bounded as $r$ approaches $r_b$. But this is not compatible with $P=P_0F^{-4}$ 
and $P(r_b)=0$. Thus we have $P(r)>0$ all over the spacetime. Since $m(r)$ is nonnegative, we also have $F(r)^\prime >0$ for $r>0$. 

For $r\rightarrow\infty$ the functions $F(r),g_{rr}(r)$ have the following asymptotic behaviour \cite{SchHo}
\begin{equation}
F(r)^2\sim \sqrt{56\pi P_0}\ r,\ \ \ \ g_{rr}(r)\sim\frac{7}{4}\ \ \ {\rm as}\  r\rightarrow\infty .\label{asymp}
\end{equation} 

\section{Entropy bound}
In this section we investigate the Bousso entropy bound for lightsheets generated by spheres. Expansion of a null geodesic 
congruence starting from a sphere orthogonally to it is $\theta =\pm 2r^{-1}EF^{-1}g_{rr}^{-1/2}$, where $+$ is for outgoing and $-$ 
for ingoing congruence and $E$ is a positive constant. Since $F>0$ and $g_{rr}>0$ everywhere, 
the lightsheet of a sphere is formed by the ingoing congruence and is ended in $r=0$. According to the definition of entropy on a 
lightlike hypersurface, the entropy on the lightsheet generated by a sphere $r=r_0,\ t=t_0$ 
is equal to entropy on a slice $r<r_0,\ t=t_0$ (we shall denote this entropy $S(r_0)$). 
Thus, in this case, the Bousso entropy bound $S_L\leq A(B)/4$ reduces to the statement 
\begin{equation}
S(r)\leq\pi r^2
\end{equation}
in Planck units. Or, defining $B(r)=S(r)/\pi r^2$, we can write $B(r)\leq 1$. From (\ref{entropy}), (\ref{sigma}) we obtain 
\begin{equation}
S(r_0)=\int_0^{r_0} 4\pi r^2\sqrt{g_{rr}(r)}\sigma _0F(r)^{-3}\D r .
\label{entropy2}
\end{equation} 
However there is another more convenient way how to express $S(r)$.
\begin{lemma} The entropy $S(r)$ of radiation contained inside a sphere of radius $r$ in a static spherically symmetric spacetime satisfies 
\begin{equation}
S(r)=\frac{\sigma _0}{24\pi P_0}\frac{F^\prime(r)}{\sqrt{g_{rr}(r)}}\ A(r),\label{Sr}
\end{equation}
where $A(r)=4\pi r^2$ is area of the sphere.
\end{lemma}
\begin{proof} For an analog of Newtonian potential $\phi ={\rm ln}F$ 
in a static perfect fluid spacetime an analog of Poisson equation holds: $\nabla _\mu\nabla ^\mu\phi =4\pi (\rho +3P)$ \cite{Beig}. 
In terms of the L-C connection $\tilde\nabla$ associated to the metric induced on a surface orthogonal to the timelike Killing field, this can be 
written as $\tilde\nabla _i\tilde\nabla ^i F=4\pi (\rho +3P)F$. Then using $\sigma =\sigma _0F^{-3},P=P_0F^{-4},\rho =3P$ we get 
\begin{equation}
\sigma =\frac{\sigma _0}{24\pi P_0}\tilde\nabla _i\tilde\nabla ^i F. 
\end{equation}
Using the Gauss theorem, 
the entropy of radiation contained in an open set $\Sigma$, which is a subset of the surface orthogonal to the Killing field, can be expressed as 
\begin{eqnarray}
S_{\Sigma}=\int _\Sigma\sigma\ {\rm vol}_\Sigma &=\frac{\sigma _0}{24\pi P_0}\int _\Sigma \tilde\nabla _i\tilde\nabla ^i F\ {\rm vol}_\Sigma =
\nonumber \\*
&=\frac{\sigma _0}{24\pi P_0}\int _{\partial\Sigma}n^i\tilde\nabla _iF\ {\rm vol}_{\partial\Sigma},
\end{eqnarray}
where $n^i$ is outer normal of $\partial\Sigma$. If $\Sigma$ is a ball of radius $r$ in spherically symmetric spacetime, 
$n^i\tilde\nabla _iF=F^\prime /\sqrt{g_{rr}}$ is constant on $\partial\Sigma$ and we get the assertion.
\end{proof}

\begin{figure}[t]
\begin{center}
\scalebox{0.37}{\rotatebox{270}{\includegraphics{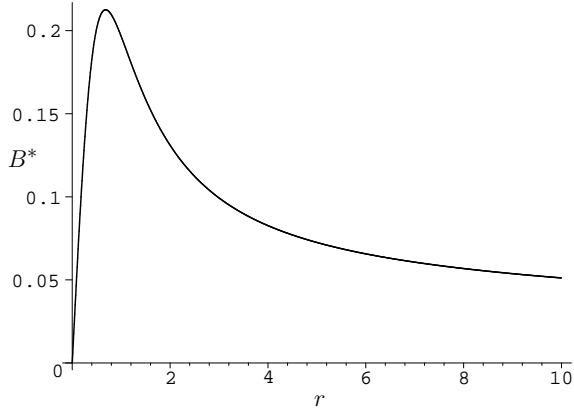}}
\put(-578,-170){\Huge$B^\ast$}
\put(-265,-415){\Huge$r$}}
\end{center}
\caption{The dependence $B^\ast(r) =S(r)/\pi r^2$ for bosons and $T_0=1,a=0,g_s=1$. \label{fig1}}
\end{figure}

Now we will show, how the function $S(r)$ depends on the parameters $T_0,a,g_s,z$. 
Choose values $T_0^\ast ,a^\ast ,g_s^\ast ,z^\ast$ of parameters $T_0,a,g_s,z$ and denote $P_0^\ast =P_0(T_0^\ast ,a^\ast ,g_s^\ast ,z^\ast)$ 
and similarly for $\sigma _0^\ast$. Next let us denote 
\[\alpha =P_0(T_0,a,g_s,z)/P_0^\ast ,\ \ \ \ \ \beta =\sigma _0(T_0,a,g_s,z)/\sigma _0^\ast .\] 
The function $S(r)$ for parameters $T_0^\ast ,a^\ast ,g_s^\ast ,z^\ast$ is 
denoted $S^\ast(r)$ and similarly for $B^\ast (r),\ g_{rr}^\ast(r)$, etc. Using (\ref{param}) and expressing $g_{rr}$ 
according to (\ref{met1}) we get $g_{rr}(r;\alpha)=g_{rr}^\ast(\sqrt{\alpha}r)$. Then using (\ref{param}) and (\ref{Sr}) we 
get 
\begin{eqnarray}
S(r ; T_0,a,g_s,z)&=&\beta\alpha ^{-3/2}S^\ast(\sqrt{\alpha}r),\label{traS}\\
B(r; T_0,a,g_s,z)&=&\beta\alpha ^{-1/2}B^\ast(\sqrt{\alpha}r).\label{traB}
\end{eqnarray}
Since $\beta\alpha ^{-1/2}$ is proportional to $T_0$
, we see that the bound $B(r)\leq 1$ can be easily violated, e.g. by choosing $T_0$ sufficiently high. To protect the 
entropy bound we have to impose a kind of Planck scale cutoff as is usual in literature \cite{Bousso1,Lemos,FMW}. 
First we will show that for $g_s$ 
small enough the entropy bound can be violated only if the temperature in the centre is of order of Planck temperature or higher. Moreover 
we will show that the violating region is always contained within a sphere of unit Planck radius for $g_s$ small enough. 

For the rest of the text we choose $T_0^\ast =1,a^\ast =0,g_s^\ast =1,z^\ast =-1$. 
The function $B^\ast(r)$ is depicted in Fig. \ref{fig1}. 
The picture was obtained numerically with help of MAPLE. Curves with arbitrary values of $T_0,a,g_s,z$ differ from the depicted 
curve only by a rescaling of the axes given by (\ref{traB}). 

\begin{figure}[t]
\begin{center}
\scalebox{0.37}{\rotatebox{270}{\includegraphics{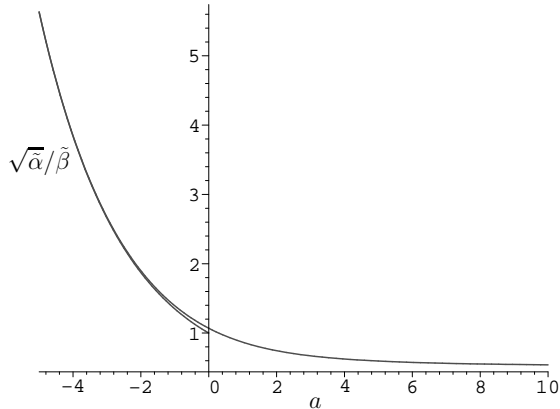}}
\put(-563,-170){\Huge$\sqrt{\tilde\alpha}/\tilde\beta$}
\put(-255,-415){\Huge$a$}}
\end{center}
\caption{The dependence $\tilde\beta ^{-1}\tilde\alpha ^{1/2}(a,z)$. For $a\leq 0$ the curve for bosons (the lower one) doesn't differ 
significantly from the curve for fermions. \label{fig2}}
\end{figure}

From the graph we see that the function $B^\ast(r)$ has a global maximum, which we denote $B^\ast _{max}$ and from 
(\ref{asymp}), (\ref{Sr}) we get that $B^\ast(r)$ decreases to zero like $r^{-1/2}$ when $r$ goes to infinity. The value of 
the maximum is approximately $B^\ast _{max}=0.2126\ ($Planck units$)$. 
To express the $T_0,g_s$ dependence of $\alpha ,\beta$ explicitly we write $\alpha =\tilde\alpha g_s T_0^4,\ \beta =\tilde\beta g_s T_0^3$, 
where $\tilde\alpha ,\tilde\beta$ depend only on $a,z$. From (\ref{traB}) we get 
\begin{equation}B_{max}(T_0,a,g_s,z)=\tilde\beta\tilde\alpha ^{-1/2}\sqrt{g_s}T_0B^\ast _{max}\label{Bmax}\end{equation} 
for a maximum of $B(r)$ for general values of the parameters. 
For given values of $a,g_s,z$ there is a critical value of temperature parameter 
$T_0=T_c$ for which $B_{max}=1$. The entropy bound is violated if and only if $T_0>T_c$. From (\ref{Bmax}) we get 
\begin{equation}T_c(a,g_s,z)=\tilde\beta ^{-1}\tilde\alpha ^{1/2}g_s^{-1/2}/B^\ast _{max}.\label{tc}\end{equation} 
Let us find a lower bound for $T_c(a,g_s,z)$. The details on $\tilde\alpha (a,z),\tilde\beta (a,z)$ are given in appendix \ref{B}. 
The function $\tilde\beta ^{-1}\tilde\alpha ^{1/2}(a,z)$ is depicted in Fig. \ref{fig2}. For $a\rightarrow -\infty$ 
the function $\tilde\beta ^{-1}\tilde\alpha ^{1/2}(a,z)$ increases exponentially in both cases $z=\pm 1$. For bosons $(z=-1,a\in(-\infty ,0])$ 
the minimal value of $\tilde\beta ^{-1}\tilde\alpha ^{1/2}(a,z)$ is $\tilde\beta ^{-1}\tilde\alpha ^{1/2}(0,-1)=1$. 
For fermions $(z=1,a\in{\rm\mathbf{R}})$ $\tilde\beta ^{-1}\tilde\alpha ^{1/2}(a,z)$ approaches $2/\sqrt{15}$ from above as $a\rightarrow\infty$. 
So we have $\tilde\beta ^{-1}\tilde\alpha ^{1/2}(a,z)>2/\sqrt{15}$ and using (\ref{tc}) we can formulate the following lemma.
\begin{lemma} The Bousso entropy bound for lightsheets generated by spheres in a static spherically symmetric spacetime filled with 
radiation is violated if and only if $T_0>T_c$, where $T_0$ is the temperature of radiation in the centre of symmetry and $T_c$, given by 
(\ref{tc}), satisfies
\begin{equation} 
T_c>\frac{2}{\sqrt{15}B^\ast _{max}}\frac{1}{\sqrt{g_s}}\approx \frac{2.429}{\sqrt{g_s}}T_{Planck}.\label{tcb}
\end{equation}
\end{lemma}
If $g_s$ is of order of unity, the temperature parameter $T_0$ (temperature 
in the centre) must be of order of Planck temperature or higher in order to get a violation of the entropy bound. 

Next we will show, that for arbitrary value of $T_0>0$, arbitrary value of $a$ (i.e. $a\leq 0$ for bosons and $a\in{\rm\mathbf{R}}$ for fermions) 
and for $g_s$ sufficiently small, a spacetime region where the bound $B(r)\leq 1$ is violated is contained in a sphere of unit Planck radius.
If the values of parameters are chosen such that there is a region where $B(r)>1$, there are 
exactly two points, where $B(r)=1$, as we can see from the Fig. \ref{fig1}. Let us denote $r_c$ the bigger value of $r$, where $B(r)=1$. The region 
violating the bound satisfies $r<r_c$. Now we will express $r_c$ as a function of the parameters and we will show that $r_c<l_P$ for $g_s$ small enough. 

Suppose that the values of parameters $T_0,a,g_s,z$ are chosen such that the region $B(r)>1$ exists. Using $B(r_c; T_0,a,g_s,z)=1$ and 
(\ref{traB}) we get 
\begin{equation} B^\ast (\sqrt{\alpha} r_c)=\sqrt{\alpha}/\beta .\label{odvozrc}\end{equation} 
Instead of $T_0$ a parameter 
$\gamma =\sqrt{\alpha}/\beta =T_0^{-1}g_s^{-1/2}\sqrt{\tilde\alpha}/\tilde\beta$ can be used. Since we suppose that a region $B(r)>1$ exists, the equation 
(\ref{odvozrc}) has to have a solution for $r_c$ and therefore $\gamma\in(0, B^\ast _{max})$. 
Denote $C(B^\ast), B^\ast\in(0,B^\ast _{max})$ an inverse function of the 
decreasing part of $B^\ast(r)$. Using (\ref{odvozrc}) we get $r_c=C(\gamma)/\sqrt{\alpha}$. 
Since $C(\gamma)$ is a one to one function, we can define a new parameter $x=C(\gamma)$ instead of $\gamma$.
The range of $x$ is $x\in(r_{max},\infty)$, where $r_{max}$ is the value of $r$, where $B^\ast(r)$ achieves its maximum. 
In terms of parameters $x,a,g_s,z$ the radius $r_c$ reads 
\begin{equation}
r_c=\sqrt{g_s}\tilde\beta ^2\tilde\alpha ^{-3/2}\ xB^\ast(x)^2.\label{rc}
\end{equation}

\begin{figure}[t]
\begin{center}
\scalebox{0.37}{\rotatebox{270}{\includegraphics{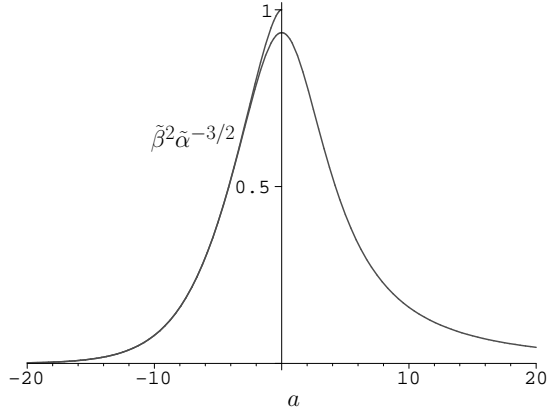}}
\put(-405,-150){\Huge$\tilde\beta ^2\tilde\alpha ^{-3/2}$}
\put(-265,-415){\Huge$a$}}
\end{center}
\caption{The dependence $\tilde\beta ^2\tilde\alpha ^{-3/2}(a,z)$. The upper curve in $a\leq 0$ region is for bosons. \label{fig3}}
\end{figure}

Let's find an upper bound for the function $r_c(x,a,g_s,z)$.  
The function $\tilde\beta ^2\tilde\alpha ^{-3/2}(a,z)$ is depicted in Fig. \ref{fig3} and we see that it has a maximum 
at $a=0$ for both $z=\pm 1$. The values of $\tilde\beta ^2\tilde\alpha ^{-3/2}(0,z)$ are $\tilde\beta ^2\tilde\alpha ^{-3/2}(0,-1)=1$ 
and $\tilde\beta ^2\tilde\alpha ^{-3/2}(0,1)=\sqrt{7/8}$, so we have 
$\tilde\beta ^2\tilde\alpha ^{-3/2}(a,z)\leq 1$.

\begin{figure}[t]
\begin{center}
\scalebox{0.37}{\rotatebox{270}{\includegraphics{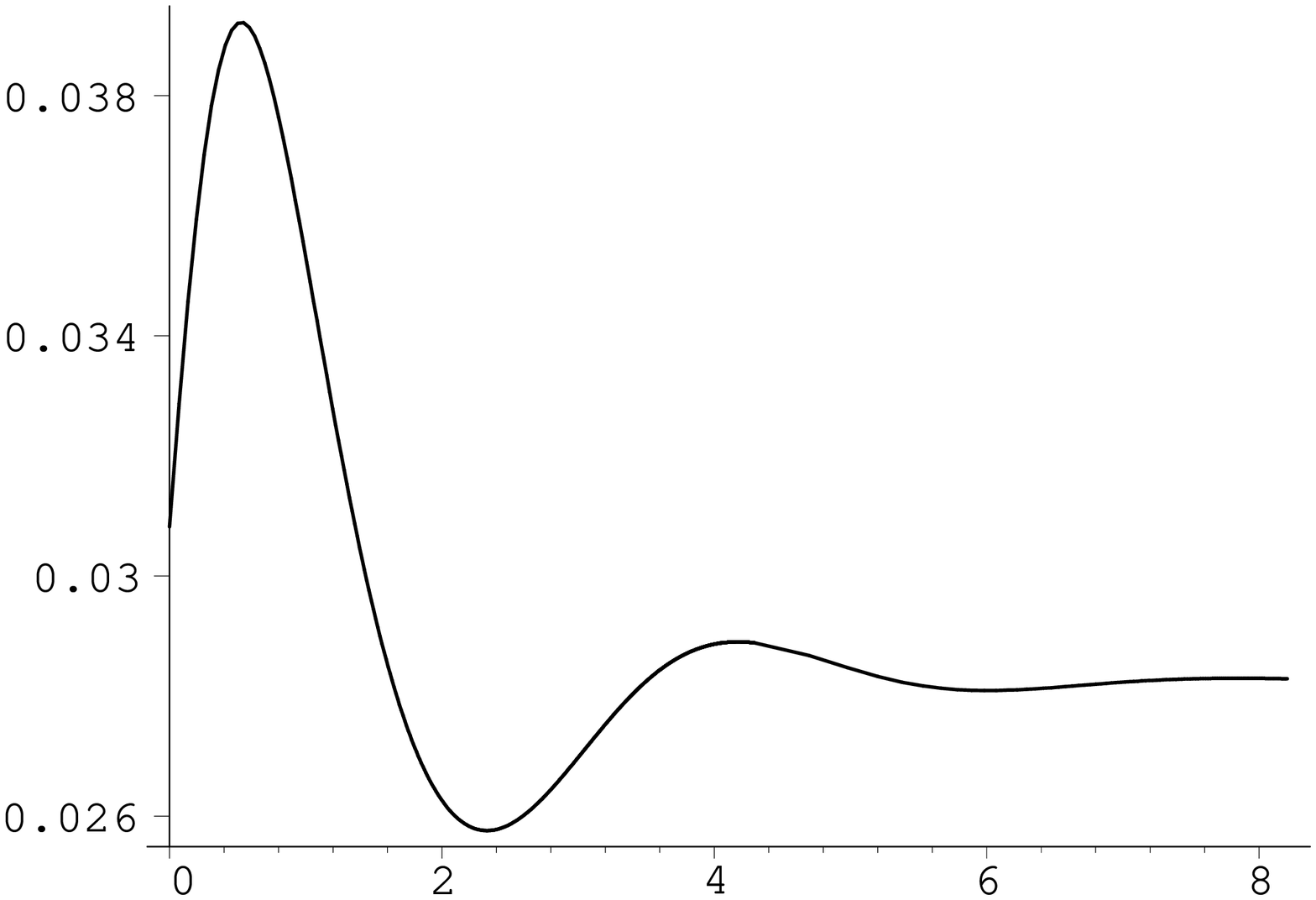}}
\put(-605,-210){\Huge$rB^{\ast 2}$}
\put(-315,-430){\Huge${\rm ln}(r/r_{max})$}}
\end{center}
\caption{The dependence $rB^\ast(r)^2$ in logarithmic scale. \label{fig4}}
\end{figure}

The dependence $xB^\ast(x)^2$ is depicted in Fig. \ref{fig4}. 
From this we see that the function $xB^\ast(x)^2$ has a global maximum and from (\ref{asymp}), (\ref{Sr}) we get that it approaches 
$(2/3)^3/\sqrt{35\pi}$ as $r$ goes to infinity. 
The value $b$ of this global maximum is approximately $b=0.0392$ (Planck units). 
Collecting all the results together we get the following conclusion.
\begin{lemma} The Bousso entropy bound for lightsheets generated by spheres in a static spherically symmetric spacetime filled with 
radiation can be violated only in a region $r<r_c$, where $r_c$, given by (\ref{rc}), satisfies 
\begin{equation} r_c\leq\sqrt{g_s}b \approx\sqrt{g_s}\ 0.0392\ l_{Planck}.\end{equation}
\end{lemma}
For $g_s$ of order not larger then $10^3$, $r_c$ will be of order not larger than unit Planck length. 
We cannot expect that physics used here for description of gas and gravity remains valid on such scales. 
So we can say that for $g_s$ small enough the entropy bound $B(r)\leq 1$ is valid on scales where classical 
physics makes sense.

\section{FMW conditions}
The sufficient conditions for Bousso entropy bound derived by Flanagan, Marolf and Wald 
(FMW conditions) \cite{FMW} read
\begin{eqnarray} 
(s_\mu k^\mu)^2&\leq &\alpha _1T_{\mu\nu}k^\mu k^\nu, \nonumber \\
\mid k^\mu k^\nu\nabla _\mu s_\nu\mid &\leq &\alpha _2T_{\mu\nu}k^\mu k^\nu
\end{eqnarray}
for an arbitrary null vector $k^\mu$, where $s^\mu=\sigma\xi ^\mu /F$ is the entropy flux and 
$\alpha _1,\alpha _2$ are arbitrary positive constants satisfying $(\pi\alpha _1)^{1/4}+(\alpha _2/\pi)^{1/2}=1$. 
In case of a stationary radiation gas these inequalities reduce to 
\begin{equation}
\frac{\sigma _0^2}{4P_0}F^{-2}\leq\alpha _1,\ \ \ \ 
\frac{\sigma _0}{P_0}\sqrt{\nabla _\mu F\nabla ^\mu F}\leq\alpha _2.\label{FMW}
\end{equation}
Comparing (\ref{FMW}) and (\ref{Sr}) and using $\nabla _\mu F\nabla ^\mu F=F^{\prime 2} /g_{rr}$ we see that 
the second FMW condition in case of a static spherically symmetric spacetime filled with radiation can be written as 
$S(r)/\pi r^2\leq\alpha _2/6\pi$. Following the lines of argumentation by which we showed that the inequality 
$S(r)/\pi r^2\leq 1$ is equivalent to $T_0\leq T_c$, where $T_c$ is given by (\ref{tc}), one can show that the second 
FMW condition is equivalent to $T_0\leq(\alpha _2/6\pi)T_c$.

Using the fact that $F(r)$ is an increasing function and $F(0)=1$, the first FMW condition can be written as $(\sigma _0^2/4P_0)\leq\alpha _1$. 
After a short calculation we get an equivalent expression of this condition: $T_0\leq T_cB^\ast _{max}(45\alpha _1/2)^{1/2}/\pi$. 

Denoting $1/\lambda$ the smaller of $B^\ast _{max}(45\alpha _1/2)^{1/2}/\pi$ and $\alpha _2/6\pi$, the set of 
conditions (\ref{FMW}) is equivalent to 
\begin{equation}T_0\leq \frac{1}{\lambda} T_c.\end{equation} 
Since $\alpha _1,\alpha _2$ are positive and satisfy 
$(\pi\alpha _1)^{1/4}+(\alpha _2/\pi)^{1/2}=1$, the smallest possible value of $\lambda$ is 
$\lambda =(\pi ^{3/4}B^{\ast -1/2} _{max}(45/2)^{-1/4}+6^{1/2})^{2}\approx 23.03$. So the restriction of temperature imposed by 
the FMW conditions is at least $23$ times more strict then restriction imposed directly by Bousso entropy bound for lightsheets 
generated by spheres. However, the FMW conditions imply the Bousso entropy bound for general lightsheets, not only for that 
generated by spheres. But even if the FMW conditions would impose stronger restrictions on parameters than the general form of the Bousso bound, it would not be surprising, since the FMW conditions are only sufficient and not necessary for the Bousso bound. Using (\ref{tcb}) we get a sufficient condition for the general form of the Bousso bound 
\begin{equation}
T_0\leq 0.105\ g_s^{-1/2}T_{Planck}.   
\end{equation}

The region where the FMW conditions are violated is always larger than the region where the Bousso bound for spherical lightsheets does not hold. This can be seen if we take the second FMW condition in the form $B(r)\leq\alpha _2/6\pi$ and we realize that the highest possible value of $\alpha _2$ is  $\alpha _2=\pi$. The range of $r$ where $B(r)>\alpha _2/6\pi$ is always wider than the range where $B(r)>1$. Let us denote $r_{FMW1}\ (r_{FMW2})$ the largest value of $r$ where the first (second) FMW condition is violated and $r_{FMW}$ the larger of the quantities $r_{FMW1}$ and $r_{FMW2}$. There is no simple relation between $r_c$ and $r_{FMW}$ like in case of temperature, but we can at least derive an upper bound for $r_{FMW}$. Repeating the argumentation which led to the formula (\ref{rc}) we obtain
\begin{eqnarray}
r_{FMW1}&=&\frac{2\pi ^2}{45\alpha _1}\sqrt{g_s}\tilde\beta ^2\tilde\alpha ^{-3/2}\ x_1F^\ast(x_1)^{-2},\\
r_{FMW2}&=&\left(\frac{6\pi}{\alpha _2}\right)^2\sqrt{g_s}\tilde\beta ^2\tilde\alpha ^{-3/2}\ x_2B^\ast(x_2)^{2} , 
\end{eqnarray}
where $x_1\in(0,\infty)$ and $x_2\in(r_{max},\infty)$ are certain parameters which are used instead of temperature. We already know that $\tilde\beta ^2\tilde\alpha ^{-3/2}\leq 1$ and $x_2B^\ast(x_2)^{2}\leq 0.0392$. Furthermore, the function $x_1F^\ast(x_1)^{-2}$ has similar behavior as the function $x_2B^\ast(x_2)^{2}$. It achieves its maximum and then it approaches a nonzero value by damped oscillations. The maximal value of $x_1F^\ast(x_1)^{-2}$ is approximately $0.2875$. Thus, we obtain the following upper bounds
\begin{eqnarray}
r_{FMW1}&\leq &\frac{2\pi ^2}{45\alpha _1}\sqrt{g_s}\ 0.2875\ l_{Planck},\\
r_{FMW2}&\leq &\left(\frac{6\pi}{\alpha _2}\right)^2\sqrt{g_s}\ 0.0392\ l_{Planck}. 
\end{eqnarray}
Denoting $\Upsilon$ the larger quantity of $0.2875\frac{2\pi ^2}{45\alpha _1}$ and $0.0392\left(\frac{6\pi}{\alpha _2}\right)^2$, the bound for $r_{FMW}$ can be written as $r_{FMW}\leq\sqrt{g_s}\ \Upsilon$. Minimizing $\Upsilon$ by appropriate choice of $\alpha _1$ and $\alpha _2$ we obtain $\Upsilon =12.58$. Thus, the final bound reads 
\begin{equation}
r_{FMW}\leq \sqrt{g_s}12.58\ l_{Planck}.
\end{equation} 
It means that for appropriate choice of $\alpha _1,\alpha _2$ (namely for values $\alpha _1=0.0100$ and $\alpha _2=1.0522$) both of the FMW conditions are satisfied for $r>\sqrt{g_s}12.58\ l_{Planck}$ no matter what are the values of parameters $T_0,a,z$.  
  
\section{Conclusion}
It was shown that the entropy $S(r)$ contained inside a sphere of radius $r$ in a static spherically symmetric spacetime filled 
with a gas of massless particles can be expressed as $S(r)=\sigma _0/(24\pi P_0)\ \dot{F}(r)\  A(r)$, where 
$A(r)=4\pi r^2$ is the area of the sphere, $F$ is the norm of a timelike Killing field, dot denotes derivative with respect to the proper radius
and $\sigma _0,P_0$ are entropy density and pressure in the centre of symmetry. 
The Bousso entropy bound for lightsheets generated by spheres was investigated and it was shown that a violation of the bound 
can occur only in a region satisfying $r<\sqrt{g_s}\ 0.0392\ l_{Planck}$, where $g_s$ is number of independent polarization states. 
It means that for $g_s$ small enough, the bound is satisfied on scales where classical physics makes sense. 
Furthermore it was shown that a central temperature of radiation $T_0$ must exceed $2.429 g_s^{-1/2} T_{Planck}$ to get a violation of the bound. 
Thus for $g_s$ small enough the bound can be protected also by a Planck temperature cutoff. If $T_0\leq 0.105\ g_s^{-1/2}T_{Planck}$ 
the Bousso bound holds for general lightsheets (not only for that generated by spheres) as was derived using the FMW conditions.
\begin{acknowledgements}
This work was partially supported by {\it Aktion Czech Republic - Austria}. I would like to thank the gravity group of Vienna University for kind 
reception. I also thank Pavel Klep\'{a}\v{c} and Prof. Jan Novotn\'{y} for helpful discussions.
\end{acknowledgements}
\appendix
\section{Proof of lemma 1\label{Z}}
From definition of $p_i$ and $\bar{p}_i$ we get $p_i=B_i\bar{p}_0+A_i^j\bar{p}_j$. 
The Jacobian of the transformation $x^i,\bar{p}_i\rightarrow x^i,p_i$ reduces 
to ${\rm det}(\partial p_i/\partial \bar{p}_j)$, which can be expressed using an identity ${\rm det}(\delta ^i_j+a^ib_j)=
1+a^ib_i$ as 
\begin{equation}
{\rm det}(\partial p_i/\partial \bar{p}_j)={\rm det}(A^j_i)(1+C_i\ \bar{p}^i/\bar{p}_0).
\label{jacobi}
\end{equation} 
Now we will show that the determinant (\ref{jacobi}) is positive and therefore 
the considered coordinate transformation preserves orientation. For the induced metric on $\Sigma$ we get 
$g_{ij}=g(\partial /\partial x^i,\partial /\partial x^j)=-B_iB_j+A^k_iA^l_j\delta _{kl}$. From this we get 
$(A^{-1})^i_k(A^{-1})^j_lg_{ij}=-C_kC_l+\delta _{kl}$. Since $\Sigma$ is spacelike, $g_{ij}$ as well as 
$(A^{-1})^i_k(A^{-1})^j_lg_{ij}$ must be positive definite. Therefore 
$(-C_kC_l+\delta _{kl})\bar{p}^k\bar{p}^l/(\bar{p}_0)^2$ must be positive. This implies 
$(C_i\ \bar{p}^i/\bar{p}_0)^2<1$. The ${\rm det}(A^j_i)$ is positive by assumption, thus all the expression (\ref{jacobi}) 
will be positive. The ${\rm det}(A^j_i)$ can be expressed as 
\begin{equation}
{\rm det}(A^j_i)=\sqrt{{\rm det}(A^k_iA^l_j\delta _{kl})}=\sqrt{{\rm det}(g_{ij}+B_iB_j)}=\sqrt{{\rm det}h_{ij}},
\end{equation} 
because $B_i=-g(\partial /\partial x^i,e_0)$ and $e_0=\xi /F$. This gives the result.
 
\section{Proof of lemma 2 \label{A}}
Consider a coordinate system $t,x^i$ with $\partial _t=\xi$ such that $\Sigma _1$ is given by $t=0$ and $\Sigma _2$ is given by $t=f(x^i)$. 
Define coordinates $x_1^i=x^i|_{\Sigma _1}$ on $\Sigma _1$ and $x_2^i=x^i|_{\Sigma _2}$ on $\Sigma _2$. Since $\Sigma _1$ and $\Sigma _2$ 
are intersected by the same set of integral lines of $\xi$, the range of $x^i_1$ and $x^i_2$ is the same. 
Further we use the following notation:   
$h_{ij}=h_{\mu\nu}(\partial /\partial x^i)^\mu (\partial /\partial x^j)^\nu ,\ 
h^1_{ij}=h_{\mu\nu}(\partial /\partial x_1^i)^\mu (\partial /\partial x_1^j)^\nu$ and similarly for $h^2_{ij}$ and $\sigma _1=\sigma |_{\Sigma _1},\ 
\sigma _2=\sigma |_{\Sigma _2}$. Now consider points $p\in\Sigma _1$ 
and $q\in\Sigma _2$ lying on the same integral line of $\xi$, i.e. $x^i_1(p)=x^i_2(q)$. Since $\partial /\partial x_2^i=
(\partial f/\partial x^i) \xi +\partial /\partial x^i,\ \partial /\partial x_1^i=\partial /\partial x^i,\ h_{\mu\nu}\xi ^\nu =0$ and $h_{\mu\nu}=h_{\nu\mu}$, 
we get $h^2_{ij}(q)=h_{ij}(q),\ h^1_{ij}(p)=h_{ij}(p)$. Since $\mathcal{L}_\xi\xi _\mu =0,\ \mathcal{L}_\xi F=0,\ \mathcal{L}_\xi g_{\mu\nu}=0$ we have 
$\mathcal{L}_\xi h_{\mu\nu}=0$ and therefore $h_{ij}$ doesn't depend on the position on an integral line of $\xi$. 
It implies that $h^1_{ij}(p)=h^2_{ij}(q)$ and therefore the functions $h^1_{ij}(x_1^i)$ and $h^2_{ij}(x^i_2)$ are the same. 
Since $\sigma =\sigma _0F^{-3}$ and $F(p)=F(q)$, also functions $\sigma _1(x^i_1)$ and $\sigma _2(x^i_2)$ are the same. Thus we get 
$S_{\Sigma _1}=S_{\Sigma _2}$. 

\section{Properties of $P_0,\sigma _0, \tilde\alpha$ and $\tilde\beta$ \label{B}}
The constants $P_0,\sigma _0$ are given by
\begin{eqnarray}
P_0 &=&\frac{g_s}{6\pi ^2}T_0^4\int^\infty _0\D q\ q^3\left({\rm e}^{q-a}+z\right)^{-1},\\
\sigma _0 &=&\frac{g_s}{2\pi ^2}T_0^3\int^\infty _0\D q\ q^2(4q/3-a)\left({\rm e}^{q-a}+z\right)^{-1}.
\end{eqnarray}
For $T_0=1,a=0,g_s=1,z=-1$ we get 
$P_0=P_0^\ast =\pi ^2/90$ and $\sigma _0=\sigma _0^\ast=4P_0^\ast$. 
The functions $\tilde\alpha =P_0/(P_0^\ast g_s T_0^4)$ and $\tilde\beta =\sigma _0/(\sigma _0^\ast g_s T_0^3)$ 
can be expressed in terms of the polylogarithm function 
\begin{eqnarray}
\tilde\alpha &=&-\frac{90}{\pi ^4}z{\rm Li}_4(-z{\rm e}^a),\label{exa}\\
\tilde\beta &=&\frac{45}{2\pi ^4}z(-4{\rm Li}_4(-z{\rm e}^a)+a{\rm Li}_3(-z{\rm e}^a)).\label{exb}\nonumber
\end{eqnarray}
For $s>1$ and $|y|\leq1$ the polylogarithm is defined as ${\rm Li}_s(y)=\sum _{n=1}^\infty y^n/n^s$. This expansion can be used in 
(\ref{exa}) for $a\leq 0$. For fermions $(z=1)$ and $a>0$ one can use another way how to express $\tilde\alpha ,\tilde\beta$ 
\begin{eqnarray}
\tilde\alpha &=&\frac{15}{4\pi ^4}a^4+\frac{15}{2\pi ^2}a^2+\frac{7}{4}+\frac{90}{\pi ^4}\ {\rm Li}_4(-{\rm e}^{-a}),\\
\tilde\beta &=&\frac{15}{4\pi ^2}a^2+\frac{7}{4}+\frac{45}{2\pi ^4}\ (4{\rm Li}_4(-{\rm e}^{-a})+a{\rm Li}_3(-{\rm e}^{-a})).\nonumber 
\end{eqnarray}

\end{document}